\documentclass[aps]{revtex4}

\usepackage{bm,epic,eepic,amsmath}
\usepackage[dvips]{graphicx}

\newcommand{\vro}{\vec\rho}
\newcommand{\vk}{\vec k}
\newcommand{\vp}{\vec p}
\newcommand{\vq}{\vec q}
\newcommand{\vx}{\vec x}
\newcommand{\vt}{\vec t}
\newcommand{\dg}{\dagger}
\newcommand{\At}{\widetilde{A}}

\begin{document}

\title{Charged Particle-Image Interaction Near a Conducting Surface}

\author{M. Durgut and E. Mete}

\affiliation{Department of Physics, Middle East Technical University, Ankara, Turkiye}

\date{\today}

\begin{abstract}
The interaction of a bulk electron with conducting surfaces is studied by means of Bohm-Pines transformation
in the second quantization formalism. The effective interaction potentials are obtained for the case of
one plane and two plane configurations in the form of electron-image electron scattering.
\end{abstract}


\maketitle

\section{Introduction}
The electron interactions in an electron gas yield collective plasmon modes. Bohm and Pines \cite{1}
developed a collective description of electrons for a three-dimensional degenerate electron gas placed in a uniform
background of positive charge. In order to obtain a plasmon description of electron interactions, a set of
supplementary field coordinates are introduced, $N'$ in number, which describe the collective motion of the system.
Hence, the model possesses a total of $3N+N'$ degrees of freedom corresponding to electrons and plasma
oscillations respectively. This extended system of electrons and plasma waves have the same
physical properties as the original system after imposing a set of $N'$ subsidiary conditions \cite{2}. Grecu \cite{3}
calculated the plasma frequency of the electron gas by treating it as a layered structure proposed by
Visscher and Falicov \cite{4} by equation of motion method in the RPA. Apostol \cite{5} pointed out that the
plasmons propagating in different layers of a solid are significantly coupled together via the electric field created by
in-plane charge for finite values of wave vector $k$. In other words, when we deal with a system of two dimensional
electron gas, the electron-plasmon coupling may not be negligible. Therefore, the quantum system consists of
electrons plus the plasmon field width the additional plasmon-plasmon coupling.
These requirements are already implemented by the canonical transformation method of Bohm and Pines. A recent attempt
employing self-energy approach to obtain the image potential near a surface is given in \cite{6}-\cite{7}.

In this paper the problem of a single bulk electron in interaction with an infinite conducting plane has been solved
for the cases of one and two planes, respectively, where the bulk electron is placed between the planes for the latter case.
A two dimensional quantum version of Bohm-Pines canonical transformation is used in second quantization formalism
(natural units are used).

The bulk electron - plane interaction effectively becomes electron-electron scattering mediated via plasmon exchanges,
where the second electron is in fact the image of the first one with respect to the conducting plane. The effective potential
evaluated for this interaction has both static and dynamic components. In the static limit the effective potential reduces
as expected to the classical potential obtained by the image method of classical electrodynamics. Similarly, the image method
result for two planes with an electron in between is also obtained.

\section{2D Quantum Version of Bohm-Pines Transformation}

The basic Hamiltonian for a two dimensional electron gas in the second quantized form is
\begin{equation}\label{eq1}
H=\int d^2\!\rho\,\psi^\dg(\vro)\,{p^2\over 2m}\, \psi(\vro)+ {1\over 2} \int d^2\!\rho\,
d^2\!\rho'\,\psi^\dg(\vro)\,\psi^\dg(\vro\,')\,V(\vro-\vro\,')\,\psi(\vro\,')\,\psi(\vro)
\end{equation}
where $\psi$ is the electron field,
\begin{equation}\label{eq2}
\psi(\vro)=\int{d^2k\over (2\pi)^2}\,e^{i\vk\cdot\vro}\,a_{\vk}
\end{equation}
and $V$ is the interparticle interaction potential. In the second quantized form $H$ becomes,
\begin{eqnarray}
\nonumber H&=&{1\over (2\pi)^4}\int d^2\!\rho\,d^2k_1\,d^2k_2{\hbar^2 k^2_2\over 2m}
e^{i(\vk_2-\vk_1)\cdot\vro}a^\dg_{\vk_1}a_{\vk_2} \\ \label{eq3} &&+{1\over 2}\int
{d^2\!\rho\,d^2\!\rho'\over (2\pi)^8}d^2k_1\,d^2k_2\,d^2k_3\,d^2k_4\,
e^{-i\vk_1\cdot\vro}e^{-i\vk_2\cdot\vro\,'}V(\vro-\vro\,')e^{i\vk_3\cdot\vro\,'}
e^{i\vk_4\cdot\vro}\,a^\dg_{\vk_1}a^\dg_{\vk_2}\,a_{\vk_3}\,a_{\vk_4}.
\end{eqnarray}
After making transformation $\vec r_1=\vro\,'$, $\vec r_2=\vro-\vro\,'$ we have,
\begin{eqnarray}
\nonumber H&=&{1\over (2\pi)^4}\int d^2k_1\,d^2k_2{\hbar^2 k^2_2\over 2m}
a^\dg_{\vk_1}a_{\vk_2}\delta(\vk_2-\vk_1) \\ \nonumber &&+{1\over 2}\int
{d^2r_1\,d^2r_2\over (2\pi)^8}d^2k_1\,d^2k_2\,d^2k_3\,d^2k_4\,a^\dg_{\vk_1}a^\dg_{\vk_2}
e^{-i\vk_1\cdot(\vec r_2+\vec r_1)}e^{-i\vk_2\cdot\vec r_1}V(r_2)e^{i\vk_3\cdot\vec r_1}
e^{i\vk_4\cdot(\vec r_2+\vec r_1)}\,a_{\vk_3}\,a_{\vk_4}\\[3mm]
\nonumber &=&{1\over (2\pi)^4}\int d^2k_1\,{\hbar^2 k^2_1\over 2m}a^\dg_{\vk_1}a_{\vk_1} \\ \label{eq4}
&&+{1\over 2}\int {d^2k_1\,d^2k_2\,d^2k_3\,d^2k_4\over (2\pi)^6}\,
\delta[(\vk_1+\vk_2)-(\vk_3+\vk_4)]\,\widetilde{V}(k_1-k_4)\,a^\dg_{\vk_1}a^\dg_{\vk_2}a_{\vk_3}\,a_{\vk_4}.
\end{eqnarray}
The substitutions $\vk_1=\vk+\vq$, $\vk_2=\vp-\vq$, $\vk_3=\vp$ and $\vk_4=\vk$ in the second term give,
\begin{equation}\label{eq5}
H=\int {d^2k\over (2\pi)^2}{\hbar^2k^2\over 2m}\,a^\dg_{\vk} a_{\vk} + \int {d^2k\over
(2\pi)^2} {d^2p\over (2\pi)^2}{d^2q\over (2\pi)^2} \,\widetilde{V}(\vq)\,
a^\dg_{\vk+\vq\,}a^\dg_{\vp-\vq\,}a_{\rule{0pt}{2.5mm}\vp\,}a_{\vk}
\end{equation}
in terms of single-particle operators. In order to obtain the collective
motion of the planar electrons, charge density operators are introduced
\begin{equation}\label{eq6}
A_{\vq}=\int {d^2k\over (2\pi)^2}\,a^\dg_{\vk+\vq}a_{\vk}
\end{equation}
with the properties
\[
\big[A_{\vq},A_{\vq'}\big]=\big[A^\dg_{\vq},A^\dg_{\vq'}\big]=\big[A_{\vq},A^\dg_{\vq'}\big]=0
\hskip2mm,\hskip5mm \big[a_{\vk},A_{\vq}\big]=a_{\vk-\vq}
\hskip2mm,\hskip5mm \big[a^\dg_{\vk},A_{\vq}\big]=-a_{\vk+\vq}\;.
\]
In terms of the density operators $A$, the second term of the Hamiltonian (\ref{eq5}) becomes,
\begin{eqnarray}
\nonumber H&=&\int {d^2k\over (2\pi)^2}\,\varepsilon_0(\vk)\,a^\dg_{\vk}a_{\vk}
+\int {d^2k\over (2\pi)^2}{d^2p\over (2\pi)^2} {d^2q\over (2\pi)^2} \,\widetilde{V}(q)
\,(-\delta_{\vp,\vk+\vq}\,a^\dg_{\vp-\vq\,}a_{\vk}+a^\dg_{\vp-\vq\,}a_{\rule{0pt}{2.5mm}\vp\,}a^\dg_{\vk+\vq\,}a_{\vk})\\[3mm]
\nonumber &=& \int {d^2k\over (2\pi)^2}\,\varepsilon_0(\vk)\,a^\dg_{\vk}a_{\vk}
-\int {d^2k\over (2\pi)^2}{d^2q\over (2\pi)^2} \,\widetilde{V}(q)\,a^\dg_{\vk}a_{\vk}
+\int {d^2k\over (2\pi)^2} \,\widetilde{V}(k) A^\dg_{\vk}A_{\vk} \\[3mm]
&=& \label{eq7} \int {d^2k\over (2\pi)^2}\,\varepsilon_0(\vk)\,a^\dg_{\vk}a_{\vk}
-\int {d^2k\over (2\pi)^2} V(0)\,a^\dg_{\vk}a_{\vk}
+\int {d^2k\over (2\pi)^2}\,\widetilde{V}(k) A^\dg_{\vk}A_{\vk}
\end{eqnarray}
where
\begin{equation}\label{eq8}
\varepsilon_0(\vk)={\vk^2\over 2m}.
\end{equation}

The potential part of the Hamiltonian (\ref{eq5}) appears to be purely kinetic term in terms of charge densities.
\begin{equation}\label{eq9}
H=\int {d^2k\over (2\pi)^2} \big[\varepsilon(\vk)a_{\vk}^\dg a_{\vk}+{1\over 2}\widetilde{V}(\vk) A_{\vk}^\dg
A_{\vk}\big]
\end{equation}
where
\begin{equation}\label{eq10}
\varepsilon(\vk)=\varepsilon_0(\vk)+V(0).
\end{equation}

The Hamiltonian in (\ref{eq9}) possesses two kinds of purely kinetic terms, one for single electrons and one for electron
densities. In order to reach the plasmon modes, Bohm and Pines\cite{2} proposed to complete this Hamiltonian by introducing
the conjugate momenta, $P_{\vk}$, such that
\begin{equation}\label{eq11}
H=H_0+H_c,
\end{equation}
where
\begin{eqnarray}
\label{eq12} H_0&=&\int {d^2k\over (2\pi)^2}\,\epsilon(k) a^\dg_{\vk} a_{\vk} \\
\label{eq13} H_c&=&{1\over 2}\int {d^2k\over
(2\pi)^2}\left[M_k^2A_{\vk}^\dg A_{\vk}+P^\dg_{\vk}P_{\vk}
+M_k\left(A_{\vk}^\dg P_{\vk}+P_{\vk}^\dg A_{\vk}\right)\right]\\
\label{eq14} M_k&=&\sqrt{\widetilde{V}(\vk)}
\end{eqnarray}
and $P_{\vk}$ satisfies,
\begin{equation}\label{eq15}
P_{\vk}\vert\Psi\rangle=0\,.
\end{equation}

The additional terms clearly do not affect the space of physical states and the subsidiary conditions (\ref{eq14})
are restrictions which turn out to be consistent with the physical properties of the system. The definition of
$M_k$ is signaling the onset of the medium-coupling region represented by plasmons. The plasmon modes are
represented by the electron density Hamiltonian, $H_c$, whereas the single electron term $H_0$ is decoupled from it.

The $k$-integration in the expression for $H_c$ is usually divided into to parts; the long range interaction defined
by the integration domain $k<k_c$ where $k_c$ is a cut-off for the collective behavior and the shorter range screened
electron interaction for $k>k_c$. However, we shall keep a $k_c$ as a cut-off parameter throughout the calculations
to explicitly demonstrate plasmon contributions, but when evaluating the effective potential, shall let $k_c\to\infty$
to include both types of interactions.

In two dimensions the Bohm-Pines transformation reads

\begin{equation}\label{eq16}
U=e^{iS}
\end{equation}
where
\begin{equation}\label{eq17}
S=\int_{k<k_c}{d^2k\over (2\pi)^2} M_k Q_{\vk} A_{\vk},
\end{equation}
$k_c$ is a physically set cut-off value for momentum and $Q_{\vk}$ are the conjugate
collective coordinates defined through
\begin{equation}\label{eq18}
\big[Q_{\vk},P_{\vk'}\big]=i(2\pi)^2\delta(\vk-\vk').
\end{equation}

The transformation of $H$ by $U$ up to second order terms in $S$,
\begin{equation}\label{eq19}
H'=U^\dg HU=H-i[S,H]-{1\over 2}[S,[S,H]]
\end{equation}
involves the following commutators :
\begin{eqnarray}
\nonumber && [S,P_{\vk}]=\int {d^2k'\over (2\pi)^2}\,M_{k'}A_{\vk'}[Q_{\vk'},P_{\vk}]=iM_kA_{\vk} \\[2mm]
&& \nonumber [S,[S,P_{\vk}]]=[S,A_{\vk}]=0 \\[2mm]
&& \nonumber [S,H_0]=\int {d^2k\over (2\pi)^2}{d^2q\over (2\pi)^2}\,
\varepsilon(\vk)M_qQ_{\vq}\,[A_{\vq},a^\dg_{\vk}a_{\vk}] \\
&& \nonumber \hskip1.2cm =\int {d^2q\over (2\pi)^2}M_qQ_{\vq} \int {d^2k\over (2\pi)^2}\,
[\varepsilon(\vk+\vq)-\varepsilon(\vk)]\,a^\dg_{\vk+\vq}a_{\vk} \\
&& \nonumber \hskip1.2cm =\int {d^2q\over (2\pi)^2}M_qQ_{\vq} \int {d^2k\over (2\pi)^2}\,
\left({\vq^{\,2}\over 2m}+{\vk\cdot\vq\over m}\right)\,a^\dg_{\vk+\vq}a_{\vk} \\[2mm]
&& \nonumber [S,[S,H_0]]=\int {d^2q_1\over (2\pi)^2}{d^2q_2\over (2\pi)^2}\,M_{q_1}M_{q_2}Q_{\vq_1}Q_{\vq_2}
\int{d^2k\over (2\pi)^2}[\varepsilon(\vk+\vq_1)-\varepsilon(\vk)]\big[A_{\vq_2},a^\dg_{\vk+\vq_1}a_{\vk}\big] \\
&& \nonumber \hskip1.7cm = \int {d^2q_1\over (2\pi)^2}{d^2q_2\over (2\pi)^2}\,M_{q_1}M_{q_2}Q_{\vq_1}Q_{\vq_2}
\int{d^2k\over (2\pi)^2}[\varepsilon(\vk+\vq_1+\vq_2)+\varepsilon(\vk)-\varepsilon(\vk+\vq_2)-\varepsilon(\vk+\vq_1)]
\,a^\dg_{\vk+\vq_1+\vq_2}a_{\vk} \\
&& \nonumber \hskip1.7cm = \int {d^2q_1\over (2\pi)^2}{d^2q_2\over (2\pi)^2}\,M_{q_1}M_{q_2}Q_{\vq_1}Q_{\vq_2}
{\vq_1\cdot\vq_2\over m}\,A_{\vq_1+\vq_2}\;.
\end{eqnarray}

In terms of the transformed conjugate momenta,
\begin{equation}\label{eq20}
P_{\vk}'= U^\dg P_{\vk}U=P_{\vk}+M_kA_{\vk}
\end{equation}
the transformed Hamiltonian, $H_c$, becomes,

\begin{eqnarray}
\label{eq21} H_c'&=& {1\over 2} \int {d^2k\over (2\pi)^2} {P'}_{\vk}^\dg P_{\vk}' \\
\label{eq22} H_0' &=& \int {d^2k\over (2\pi)^2} \epsilon(\vk) a_{\vk}^\dg a_{\vk} + i
\int {d^2q\over (2\pi)^2} M_q Q_{\vq} \int {d^2k\over (2\pi)^2} {\vq\cdot\left( \vk+{\vq\over 2}\right)\over m}\,
a_{\vk+\vq}^\dg a_{\vk}  \\ \nonumber && - \int {d^2q_1\over (2\pi)^2}
{d^2q_2\over (2\pi)^2} M_{q_1} M_{q_2} Q_{\vq_1} Q_{\vq_2} {\vq_1\cdot\vq_2\over m} A_{\vq_1+\vq_2}\,.
\end{eqnarray}

If the term with $\vq_1=-\vq_2$ in the third integral is singled out and then $Q_{-\vq}=-Q^\dg_{\vq}$ is used,
the Hamiltonian for the system turns out to be

\begin{eqnarray}
\nonumber H'&=&\int {d^2k\over (2\pi)^2}\,\varepsilon(\vk)a^\dg_{\vk}a_{\vk}
+i\int {d^2q\over (2\pi)^2}{d^2k\over (2\pi)^2}M_qQ_{\vq}{\vq\cdot\left(\vk+{\vq\over 2}\right)\over m}
\,a^\dg_{\vk+\vq}a_{\vk} \\[1mm] \nonumber &&
+{1\over 2}\int_{\vq_1\neq\vq_2} {d^2k\over (2\pi)^2}{d^2q_1\over (2\pi)^2}
{d^2q_2\over (2\pi)^2}\,M_{q_1}M_{q_2}Q^\dg_{\vq_1}Q_{\vq_2}{\vq_1\cdot\vq_2\over m}a^\dg_{\vk-\vq_1+\vq_2}a_{\vk}\\[1mm]
\label{eq23} && +\int {d^2k\over (2\pi)^2}\,M_k^2Q^\dg_{\vk}Q_{\vk}{k^2\over m}\int {d^2k'\over (2\pi)^2}\,a^\dg_{\vk'}a_{\vk'}
+{1\over 2}\int {d^2k\over (2\pi)^2}\,P_{\vk}^\dg P_{\vk}\,.
\end{eqnarray}

Finally the total Hamiltonian can be represented as,

\begin{equation}\label{eq24}
H'=H_0+H_{\rm pl}+H_{\rm el-pl}
\end{equation}
where

\begin{eqnarray}
\label{eq25} H_0 &=& \int {d^2k\over (2\pi)^2} \varepsilon(\vk) a_{\vk}^\dg a_{\vk}
\hspace{4mm};\hspace{4mm}\varepsilon(\vk)={\vk^{\,2}
\over 2m}-{1\over 2}V(0), \\ \noalign{\vskip3pt}
\label{eq26} H_{\rm pl} &=& {1\over 2} \int {d^2k\over (2\pi)^2} \Big[ P_{\vk}^\dg P_{\vk}
+ {k^2M_k^2\over m} Q_{\vk}^\dg Q_{\vk} \hat n \Big] \hspace{4mm};\hspace{4mm}
\hat n=a_{\vk}^\dg a_{\vk}, \\ \noalign{\vskip3pt} \nonumber
H_{\rm el-pl} &=& i\int {d^2k\over (2\pi)^2} {d^2q\over (2\pi)^2} M_qQ_{\vq}
{\vq\cdot\big(\vk+{\vq\over 2}\big)\over m} a_{\vk}^\dg a_{\vk} \\ \noalign{\vskip2pt}
\label{eq27} && + {1\over 2} \int_{\vq+1\neq\vq_2} {d^2k\over (2\pi)^2}
{d^2q_1\over (2\pi)^2}\, {d^2q_2\over (2\pi)^2} M_{q_1} M_{q_2}
Q_{\vq_1}^\dg Q_{\vq_2} {\vq_1\cdot\vq_2\over m}\,
a_{\vk-\vq_1+\vq_2}^\dg a_{\vk}.
\end{eqnarray}

$H_0$ is the free energy of single electrons, $H_{\rm pl}$ represents the quantization
of planar charge density oscillations (plasmons) with momenta $P$ and position $Q$. The electron-plasmon interaction,
$H_{\rm el-pl}$ expresses the interaction of single electrons with plasmons.

A comment on the subsidiary conditions introduced in (\ref{eq15}) is in order here. After the transformation
constraint (\ref{eq15}) becomes
\begin{equation}\label{eq28}
\Big(P'_{\vk}-M_kA_{\vk}\Big)\vert\Psi\rangle=0
\end{equation}
where $A_{\vk}$ corresponds to the surface charge density $\sigma$ and $M_k$ is given in Eq.\ref{eq14}.
Eq.\ref{eq28} is actually the Gauss' Law,
\begin{equation}\label{eq29}
kP'_{\vk}=4\pi A_{\vk}\hskip5mm\Rightarrow\hskip5mm \nabla\cdot\vec E=4\pi\rho
\end{equation}
where $\vec E$ is the electric field due to planar charge and $\rho$ is the charge density. Therefore, Bohm and Pines
procedure incorporates new degrees of freedom without disturbing the known physics.

\section{Effective Electron-Plane Interaction}

The effective potential for an infinite conducting plane - single bulk charged particle
interaction can be evaluated by using the two dimensional Bohm-Pines transformation given
in the previous section. A single electron is located a distance $|\vec z|=d$ away from an infinite
electrically neutral metallic sheet. In-plane electron gas is in a uniform background of positive
charge, therefore, the electrons can be assumed to be quasi-free particles.

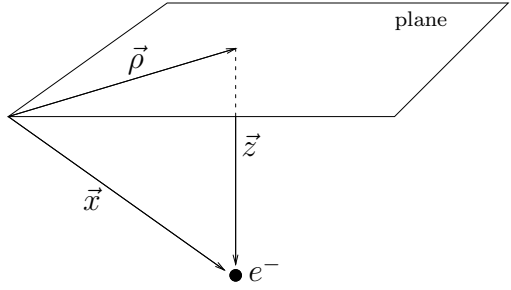
\begin{figure}[ht]
\setlength{\unitlength}{0.0004in}
\begingroup\makeatletter\ifx\SetFigFont\undefined%
\gdef\SetFigFont#1#2#3#4#5{%
  \reset@font\fontsize{#1}{#2pt}%
  \fontfamily{#3}\fontseries{#4}\fontshape{#5}%
  \selectfont}%
\fi\endgroup%
{\renewcommand{\dashlinestretch}{30}
\begin{picture}(6624,4200)(0,-150)
\put(3012,83){\blacken\ellipse{150}{150}}
\put(3012,83){\ellipse{150}{150}}
\path(2112,3683)(6612,3683)(5112,2183)
    (12,2183)(2112,3683)(2112,3683)
\path(3012,2183)(3012,233)
\path(3012,2183)(3012,233)
\path(2982.000,353.000)(3012.000,233.000)(3042.000,353.000)
\path(12,2183)(3012,3083)
\path(12,2183)(3012,3083)
\path(2905.681,3019.783)(3012.000,3083.000)(2888.440,3077.253)
\dashline{60.000}(3012,3083)(3012,2183)
\path(12,2183)(2862,158)
\path(12,2183)(2862,158)
\path(2746.802,203.049)(2862.000,158.000)(2781.555,251.960)
\put(3180,8){\makebox(0,0)[lb]{\smash{{{\SetFigFont{12}{14.4}{\rmdefault}{\mddefault}{\updefault}$e^-$}}}}}
\put(5112,3383){\makebox(0,0)[lb]{\smash{{{\SetFigFont{12}{14.4}{\rmdefault}{\mddefault}{\updefault}\footnotesize{plane}}}}}}
\put(3087,1658){\makebox(0,0)[lb]{\smash{{{\SetFigFont{12}{14.4}{\rmdefault}{\mddefault}{\updefault}$\vec z$}}}}}
\put(1587,2825){\makebox(0,0)[lb]{\smash{{{\SetFigFont{12}{14.4}{\rmdefault}{\mddefault}{\updefault}$\vro$}}}}}
\put(1000,950){\makebox(0,0)[lb]{\smash{{{\SetFigFont{12}{14.4}{\rmdefault}{\mddefault}{\updefault}$\vx$}}}}}
\end{picture}
}
\caption{Bulk electron-plane configuration}
\end{figure}

The interaction Hamiltonian for the system is
\begin{equation}\label{eq30}
H_I=\int dz\,d^2\!\rho\, d^2\!\rho'\,V(\vro-\vro\,',z)\phi^\dg(\vec x)\psi^\dg(\vro\,')
\psi(\vro\,')\phi(\vec x),
\end{equation}
where $\psi(\vro)$ corresponds to the electron field as given in Eq.\ref{eq2} extending over the plane, and
$\phi(\vec x)$ is the electron field defined in the bulk defined as
\begin{equation}\label{eq31}
\phi(\vx)=\int {d^3p\over (2\pi)^3}\,e^{i\vp\cdot\vx}b_{\vp}\,.
\end{equation}
We let momentum vector $\vk$ be conjugate to position vector $\vro$ in two dimensions. Hence, in three dimensional
space, momentum vector $\vp=\vk+\ell\hat k$ will be conjugate to position vector $\vec x=\vro+z\hat k$.

Since Coulomb interaction depends on interparticle distances, it can be expressed in
terms of momentum flows in momentum space.
\begin{eqnarray}
\nonumber H_I&=& \int d^2\!\rho\, d^2\!\rho' {d^3p_1\over (2\pi)^3}{d^2k'_1\over (2\pi)^2}
{d^2k'_2\over (2\pi)^2}{d^3p_2\over (2\pi)^3}\,e^{i(\vk_2-\vk_1)\cdot\vro}e^{i(\vk'_2-\vk'_1)\cdot\vro\,'}
\left[\int dz\, e^{i(\ell_2-\ell_1)z}\,V(\vro-\vro\,',z)\right]b^\dg_{\vp_1}a^\dg_{\vk'_1}a_{\vk'_2}b_{\vp_2}\\[1mm]
\nonumber &=& \int {d^3p_1\over (2\pi)^3}{d^2k'_1\over (2\pi)^2}{d^2k'_2\over (2\pi)^2}{d^3p_2\over (2\pi)^3}\,
\left[\int d^2\!\rho\, d^2\!\rho'\,e^{i(\vk_2-\vk_1)\cdot\vro}e^{i(\vk'_2-\vk'_1)\cdot\vro\,'}
\widetilde{V}(\vro-\vro\,',\ell_2-\ell_1)\right]b^\dg_{\vp_1}a^\dg_{\vk'_1}a_{\vk'_2}b_{\vp_2}\,.\\[3mm]
\noalign{\rm Substitution $\vec r_1=\vro\,'$ and $\vec r_2=\vro-\vro\,'$ yields,}\nonumber \\[1mm]
\nonumber &=& \int {d^3p_1\over (2\pi)^3}{d^2k'_1\over (2\pi)^2}{d^2k'_2\over (2\pi)^2}{d^3p_2\over (2\pi)^3}
\left[\int d^2r_2\,\widetilde{V}(r_2,\ell_2-\ell_1)e^{i(\vk_2-\vk_1)\cdot\vec r_2}
\int d^2r_1\,e^{i[(\vk_2-\vk_1)+(\vk'_2-\vk'_1)]\cdot\vec r_1}\right]b^\dg_{\vp_1}a^\dg_{\vk'_1}a_{\vk'_2}b_{\vp_2}\\[1mm]
\nonumber &=& \int {d^3p_1\over (2\pi)^3}{d^3p_2\over (2\pi)^3}{d^2k'_1\over (2\pi)^2}\,d^2k'_2\,
U(\vk_2-\vk_1,\ell_2-\ell_1)\delta(\vk_2-\vk_1+\vk\,'_2-\vk\,'_1)\,b^\dg_{\vp_1}a^\dg_{\vk_1'}a_{\vk_2'}b_{\vp_2}\\[1mm]
\nonumber &=& \int {d^3p_1\over (2\pi)^3}{d^3p_2\over (2\pi)^3}{d^2k'_1\over (2\pi)^2}\,d^2k'_2\,
U(\vp_2-\vp_1)\delta(\vk_2-\vk_1+\vk\,'_2-\vk\,'_1)\,b^\dg_{\vp_1}b_{\vp_2}a^\dg_{\vk_1'}a_{\vk_2'}\\[1mm]
\nonumber &=& \int {d^3p_1\over (2\pi)^3}{d^3p_2\over (2\pi)^3}{d^2k'_1\over (2\pi)^2}
U(\vp_2-\vp_1)\,b^\dg_{\vp_1}b_{\vp_2}a^\dg_{\vk_1'}a_{\vk_1+\vk_1'-\vk_2}\,.\\[3mm]
\noalign{\rm Another substitution $\vp\,'=\vp_2$ and $\vp=\vp_2-\vp_1$ gives the interaction term,}\nonumber \\[1mm]
\label{eq32} &=&\int {d^2k\over (2\pi)^2} {d^3p\over (2\pi)^3} {d^3p'\over (2\pi)^3}
\,U(\vp)\,b^\dg_{\vp\,'-\vp}\,b^{}_{\vp\,'}\,a^\dg_{\vk+\vk_2-\vk_1}\,a^{}_{\vk},
\end{eqnarray}
where
\begin{equation}\label{eq33}
U(\vp)=U(\vk_2-\vk_1,\ell_2-\ell_1)=(2\pi)^2\int d^2\!\rho
\;\widetilde{V}(\vro,\ell_2-\ell_1)\,e^{i(\vk_2-\vk_1)\cdot\vro}
\end{equation}
and
\begin{equation}\label{eq34}
\widetilde{V}(\vro,\ell_2-\ell_1)=\int dz\,e^{i(k_{2_z}-k_{1_z})z}\,V(\vro,z)\;.
\end{equation}

Momentum space vector $\vk_2-\vk_1$ conjugate to the position vector $\vro$ which lies
over the surface, can be considered as the parallel component of the vector $\vp$ :
$\vk_2-\vk_1=\vp_\parallel$. This interpretation is immediately apparent from the
Eq.\ref{eq32} that there is no conservation for the perpendicular component
of incoming momentum $\vp$. This is because the metallic sheet is assumed to be rigid.

In terms of charge densities the interaction Hamiltonian, $H_I$, can be
written as,
\begin{equation}\label{eq35}
H_I=\int {d^3p\over (2\pi)^3}\,U(\vp)\,B^\dg_{\vp}A^{}_{\vp_\parallel}
\end{equation}
where
\begin{eqnarray}
\label{eq36} B_{\vp}^\dg &=& \int {d^3p'\over (2\pi)^3}\,b_{\vp-\vp\,'}^\dg b^{}_{\vp\,'} \\
\label{eq37} A_{\vp_\parallel} &=& \int {d^2k\over (2\pi)^2}\,a^\dg_{\vk+\vp_\parallel} a^{}_{\vk}\;.
\end{eqnarray}

This interaction term can be rewritten by separating the momentum vector $\vp$ to
its parallel and perpendicular components :
\begin{equation}\label{eq38}
H_I=\int {d^2k\over (2\pi)^2}\,m_{\vk}^\dg A_{\vk},
\end{equation}
where
\begin{equation}\label{eq39}
m_{\vk}^\dg=\int {dp_{\perp}\over 2\pi}\,U(\vk,\vp_\perp)B_{\vk,\vp_\perp}^\dg\;.
\end{equation}

The new Hamiltonian for the system consists of $H'$, the Hamiltonian in the absence of bulk electron
given in Eq.\ref{eq9}, free energy of the bulk electron, $H_{0,{\rm bulk}}$, and its interaction with the plane,
\begin{eqnarray}
\nonumber H_{\rm new}&=&H'+H_{0,{\rm bulk}}+H_I\\
\nonumber &=& H_0+H_{0,{\rm bulk}}+{1\over 2}\int {d^2k\over (2\pi)^2}\big[M_k^2A^\dg_{\vk}A_{\vk}+2m_{\vk}^\dg A_{\vk}\big]\\
\label{eq40} &=&H_0+H_{0,{\rm bulk}}+{1\over 2}\int {d^2k\over (2\pi)^2}\Big[M_k^2{A'}_{\vk}^\dg {A'}^{}_{\vk}
-{m^\dg_{\vk}m_{\vk}^{}\over M_k^2}\Big],
\end{eqnarray}
where
\begin{equation}\label{eq41}
{\At}_{\vk}=A_{\vk}+{m_{\vk}\over M_k^2}.
\end{equation}
In order to employ the Bohm-Pines transformation we complete the Hamiltonian, $H$, by
adding the plasmon-bulk coupling terms.
\begin{equation}\label{eq42}
H_{c,{\rm new}}={1\over 2}\int {d^2k\over (2\pi)^2}\Big[M_k^2{\At}_{\vk}^\dg {\At}^{}_{\vk}
-{m^\dg_{\vk}m_{\vk}^{}\over M_k^2}+P_{\vk}^\dg P^{}_{\vk}
+M_k\Big({\At}^\dg_{\vk}P_{\vk}+P^\dg_{\vk}{\At}_{\vk}\Big)\Big].
\end{equation}
The Bohm-Pines transformation for $H_{c,{\rm new}}$ yields the familiar results,
\begin{eqnarray}
\label{eq43} P'_{\vk}&=&P_{\vk}+M_k{\At}_{\vk} \\
\label{eq44} H'_{c,{\rm new}}&=&{1\over 2} \int {d^2k\over (2\pi)^2}
\Big({P'}_{\vk}^\dg {P'}^{}_{\vk}-{m^\dg_{\vk}m_{\vk}^{}\over M_k^2}\Big).
\end{eqnarray}
The second term in the transformed Hamiltonian $H'_{c,{\rm new}}$ is the effective interaction Hamiltonian for the
bulk electron - plane system. With the definitions (\ref{eq36}) and (\ref{eq39}), the effective Hamiltonian,
$H_{\rm eff}$, can be written as,
\begin{subequations}
\begin{eqnarray}
\label{eq45a} H_{\rm eff} &=& -{1\over 2}\int {d^2k\over (2\pi)^2}{m^\dg_{\vk}m_{\vk}\over M_k^2}\\[1mm]
\label{eq45b} &=& -{1\over 2}\int {d^2k\over (2\pi)^2}{d\ell\over 2\pi} {d\ell'\over 2\pi}
{U(\vk,\ell)U(\vk,\ell')\over\widetilde{V}(\vk)}\,B_{\vk,\ell}^\dg B_{\vk,\ell'}\\[1mm]
\label{eq45c} &=&-{1\over 2}\int_{k<k_c} {d^3p_1\over (2\pi)^3}{d^3p_2\over (2\pi)^3}{d^2k\over (2\pi)^2}
{d\ell\over 2\pi} {d\ell'\over 2\pi} {U(\vk,\ell)U(\vk,\ell')\over\widetilde{V}(\vk)}\,b^\dg_{\vp_1-\vp\,'}
b^{}_{\vp_1} b^\dg_{\vp_2+\vp\,}b^{}_{\vp_2}
\end{eqnarray}
\end{subequations}
where $\vp=(\vk,\ell)$, $\vp\,'=(\vk,\ell')$.
This is nothing but the scattering of two electrons effectively replacing the original electron - plane interaction.
In terms of three dimensional particle momenta $\vp_1$ and $\vp_2$, the scattering diagram is shown in Fig.2.

\begin{figure}[ht]
\setlength{\unitlength}{0.0004in}
\begingroup\makeatletter\ifx\SetFigFont\undefined%
\gdef\SetFigFont#1#2#3#4#5{%
  \reset@font\fontsize{#1}{#2pt}%
  \fontfamily{#3}\fontseries{#4}\fontshape{#5}%
  \selectfont}%
\fi\endgroup%
{\renewcommand{\dashlinestretch}{30}
\begin{picture}(5424,3539)(0,-200)
\path(2712,2112)(2712,912)
\path(2712,2112)(2712,912)
\dottedline{45}(12,1512)(5412,1512)
\path(912,3012)(2712,2112)
\path(912,3012)(2712,2112)
\path(2712,912)(912,12)
\path(2712,912)(912,12)
\path(2712,912)(4512,12)
\path(2712,912)(4512,12)
\path(2712,2112)(4512,3012)
\path(2712,2112)(4512,3012)
\path(912,3012)(1812,2562)
\path(912,3012)(1812,2562)
\path(1691.252,2588.833)(1812.000,2562.000)(1718.085,2642.498)
\path(912,12)(1812,462)
\path(912,12)(1812,462)
\path(1718.085,381.502)(1812.000,462.000)(1691.252,435.167)
\path(2712,2112)(3612,2562)
\path(2712,2112)(3612,2562)
\path(3518.085,2481.502)(3612.000,2562.000)(3491.252,2535.167)
\path(2712,912)(3612,462)
\path(2712,912)(3612,462)
\path(3491.252,488.833)(3612.000,462.000)(3518.085,542.498)
\path(2712,912)(2712,1437)
\path(2712,912)(2712,1437)
\path(2742.000,1317.000)(2712.000,1437.000)(2682.000,1317.000)
\put(1200,2350){\makebox(0,0)[lb]{\smash{{{\SetFigFont{12}{14.4}{\rmdefault}{\mddefault}{\updefault}$\vp_2$}}}}}
\put(3800,2350){\makebox(0,0)[lb]{\smash{{{\SetFigFont{12}{14.4}{\rmdefault}{\mddefault}{\updefault}$\vp_2+(\vk,\ell)$}}}}}
\put(1200,550){\makebox(0,0)[lb]{\smash{{{\SetFigFont{12}{14.4}{\rmdefault}{\mddefault}{\updefault}$\vp_1$}}}}}
\put(3800,550){\makebox(0,0)[lb]{\smash{{{\SetFigFont{12}{14.4}{\rmdefault}{\mddefault}{\updefault}$\vp_1-(\vk,\ell')$}}}}}
\put(2326,1050){\makebox(0,0)[lb]{\smash{{{\SetFigFont{12}{14.4}{\rmdefault}{\mddefault}{\updefault}$\vk$}}}}}
\end{picture}
}
\caption{Electron scattering by the surface}
\end{figure}
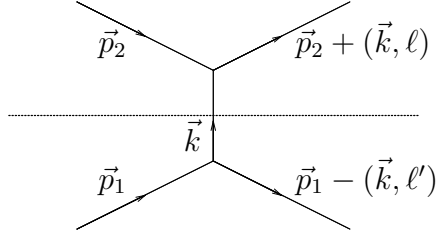

In terms of field operators this term can be written as,
\begin{equation}\label{eq46}
H_{\rm eff}={-1\over 2}\int d^3\vec x\,d^3\vec x'\,\phi^\dg(\vec x)\phi^\dg(\vec x')
\left\{\int_{k<k_c} {d^2k\over (2\pi)^2}{d\ell\over 2\pi}{d\ell'\over 2\pi}
{U(\vk,\ell)U(\vk,\ell')\over \widetilde{V}(\vk)}\,e^{i\vk\cdot(\vro-\vro\,')+i\ell'z+
i\ell z'}\right\}\phi(\vec x')\phi(\vec x).
\end{equation}
The characterization of $H_{\rm eff}$ as an effective interaction in three dimensions by the three dimensional fields
$\phi(\vx)$ is analogous to the replacement of classical ``charge-conducting plane" problem to an equivalent ``charge-image
charge" problem.

The integral inside the curly brackets in equation (\ref{eq46}) is an effective two-particle interaction potential.
By introducing the Coulomb potential, $U(\vec k,\ell)=4\pi e^2/(k^2+\ell^2)$
and $\widetilde{V}(\vk)=2\pi e^2/k$, one can calculate the effective potential as,
\begin{subequations}
\begin{eqnarray}
\label{eq47a} V_{\rm eff}(\vec x,\vec x',k_c)&=&-\int {d^2k\over (2\pi)^2}{e^{i\vk\cdot(\vro-\vro\,')}\over \widetilde{V}(\vk)}
\int{d\ell\over (2\pi)}\,U(\vk,\ell)e^{i\ell z'}\int {d\ell'\over (2\pi)}\,U(\vk,\ell')e^{-i\ell' z}\\[1mm]
\nonumber &=& -\int {d^2k\over (2\pi)^2}{e^{i\vk\cdot(\vro-\vro\,')}\over \widetilde{V}(\vk)}
\left(4\pi e^2\int {d\ell'\over 2\pi} {e^{i\ell'z}\over k^2+{\ell'}^2}\right)
\left(4\pi e^2\int {d\ell\over 2\pi} {e^{-i\ell z'}\over k^2+\ell^2}\right)\\[1mm]
\nonumber &=& -\int {d^2k\over (2\pi)^2}{e^{i\vk\cdot(\vro-\vro\,')}\over \widetilde{V}(\vk)}
\left({2\pi e^2\over k}e^{-kz}\right)
\left({2\pi e^2\over k}e^{-kz'}\right)\\[1mm]
\nonumber &=& -\int {d^2k\over (2\pi)^2}e^{i\vk\cdot(\vro-\vro\,')}e^{-k(z+z')}\widetilde{V}(\vk) \\[1mm]
\nonumber &=& -e^2\int_0^{2\pi}{d\theta\over 2\pi} \int_0^{k_c}dk\,e^{k(i\vert\vro-\vro\,'\vert\cos\theta-z-z')} \\[1mm]
\label{eq47b} &=&-{ie^2\over \pi b}\int_0^\pi {d\theta\over \cos\theta+i{a\over b}}\,
\left[1-e^{k_c(ib\cos\theta-a)}\right]
\end{eqnarray}
\end{subequations}
where $a=z+z'$ and $b=|\vro-\vro\,'|$ for simplicity.

The first term of the integral in Eq.\ref{eq47b} is the dominant contribution to $V_{\rm eff}$. The second term becomes
negligible for large $a$ or large $b$ when $k_c$ is fixed, or for large $k_c$ with $a,b$ fixed. In the $k_c\to\infty$ limit,
the result becomes,
\begin{eqnarray}
\nonumber V_{\rm eff}(\vec x,\vec x')&=& -{i\over\pi}{e^2\over b}\int_0^\pi {d\theta\over \cos\theta+i{a\over b}} \\[1mm]
\nonumber &=& -{i\over\pi}{e^2\over b}\left({-i\pi\over \sqrt{1+{a^2\over b^2}}}\right) \\[1mm]
\label{eq48}&=&{-e^2\over \sqrt{|\vro-\vro\,'|^2+(z+z')^2}}.
\end{eqnarray}
Contact with the classical image method result is made for $\vx=\vx\,'$, in which case $V_{\rm eff}=-e^2/2z$.

\section{Two Planes}

\begin{figure}[ht]
\setlength{\unitlength}{0.0004in}
\begingroup\makeatletter\ifx\SetFigFont\undefined%
\gdef\SetFigFont#1#2#3#4#5{%
  \reset@font\fontsize{#1}{#2pt}%
  \fontfamily{#3}\fontseries{#4}\fontshape{#5}%
  \selectfont}%
\fi\endgroup%
{\renewcommand{\dashlinestretch}{30}
\begin{picture}(4857,4239)(0,-10)
\put(2430,1812){\blacken\ellipse{150}{150}}
\put(2430,1812){\ellipse{150}{150}}
\path(30,3012)(1530,4212)
\path(1530,4212)(5830,4212)
\path(30,3012)(4330,3012)
\path(4330,3012)(5830,4212)
\path(30,3012)(2190,1962)
\blacken\thicklines
\path(2068.960,1987.482)(2190.000,1962.000)(2095.192,2041.444)(2068.960,1987.482)
\thinlines
\path(30,3012)(2430,3612)
\blacken\thicklines
\path(2320.859,3553.791)(2430.000,3612.000)(2306.307,3612.000)(2320.859,3553.791)
\thinlines
\path(2430,3012)(2430,2112)
\blacken\thicklines
\path(2400.000,2232.000)(2430.000,2112.000)(2460.000,2232.000)(2400.000,2232.000)
\thinlines
\path(30,12)(4330,12)
\path(12,16)(1512,1216)
\path(4332,14)(5832,1214)
\path(1545,1212)(5845,1212)
\blacken\thicklines
\path(2460.000,1392.000)(2430.000,1512.000)(2400.000,1392.000)(2460.000,1392.000)
\path(2430,1512)(2430,612)
\thinlines
\path(30,3012)(2430,612)
\blacken\thicklines
\path(2323.934,675.640)(2430.000,612.000)(2366.360,718.066)(2323.934,675.640)
\thinlines
\path(30,12)(2430,612)
\blacken\thicklines
\path(2320.859,553.791)(2430.000,612.000)(2306.307,612.000)(2320.859,553.791)
\thinlines
\dottedline{45}(2430,3612)(2430,3012)
\put(1300,3760){\makebox(0,0)[lb]{\smash{{{\SetFigFont{12}{14.4}{\rmdefault}{\mddefault}{\updefault}$\vec r_1\!=\!\vro_1$}}}}}
\put(2580,2562){\makebox(0,0)[lb]{\smash{{{\SetFigFont{12}{14.4}{\rmdefault}{\mddefault}{\updefault}$\vec z$}}}}}
\put(1455,2487){\makebox(0,0)[lb]{\smash{{{\SetFigFont{12}{14.4}{\rmdefault}{\mddefault}{\updefault}$\vx$}}}}}
\put(780,1662){\makebox(0,0)[lb]{\smash{{{\SetFigFont{12}{14.4}{\rmdefault}{\mddefault}{\updefault}$\vec r_2$}}}}}
\put(1230,537){\makebox(0,0)[lb]{\smash{{{\SetFigFont{12}{14.4}{\rmdefault}{\mddefault}{\updefault}$\vro_2$}}}}}
\put(2630,1812){\makebox(0,0)[lb]{\smash{{{\SetFigFont{12}{14.4}{\rmdefault}{\mddefault}{\updefault}$e^-$}}}}}
\put(2520,770){\makebox(0,0)[lb]{\smash{{{\SetFigFont{12}{14.4}{\rmdefault}{\mddefault}{\updefault}-$\vec z\!+\!d\hat k$}}}}}
\put(4150,3900){\makebox(0,0)[lb]{\smash{{{\SetFigFont{12}{14.4}{\rmdefault}{\mddefault}{\updefault}\footnotesize plane 1}}}}}
\put(4150,900){\makebox(0,0)[lb]{\smash{{{\SetFigFont{12}{14.4}{\rmdefault}{\mddefault}{\updefault}\footnotesize plane 2}}}}}
\end{picture}
}
\caption{Plane-bulk electron-plane configuration}
\label{fig3}
\end{figure}
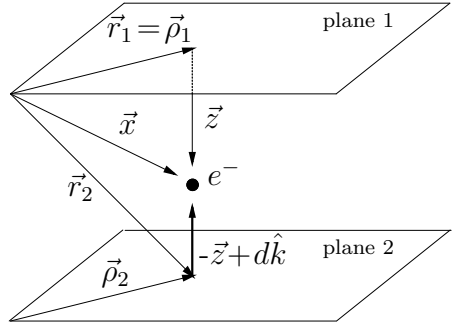

In this case, a single bulk electron placed in between two infinite neutral conducting planes which have a separation $d$ between
them. The configuration of the problem is shown in Fig.\ref{fig3}. The planes act as 2 dimensional electron gas structures which
interact with a bulk electron. $\vx=\vec r_1+\vec z\hat k=\vec r_2-\vec z+d\hat k$ denotes the position of the bulk electron
and is conjugate to $\vp=\vp_\parallel+\vp_\perp=\vk+\ell\hat k$.

The Hamiltonian for this system involves plane-plane and plane-bulk electron interactions
as well as the free energies of surface charges and the bulk electron.
\begin{equation}\label{eq49}
H=H_0^{(1)}+H_0^{(2)}+H_0^{({\rm v})}+H_{\rm ee}^{(1)}+H_{\rm ee}^{(2)}+H_{\rm pv}^{(1)}+H_{\rm pv}^{(2)}+H_{\rm pp}
\end{equation}
where $H^{(1)}_0$, $H^{(2)}_0$ represents free energies of the plane electrons, $H_0^{({\rm v})}$ is the free energy of the
bulk electron, $H_{ee}$ is the interaction of electrons within the planes or
it comes out as a kinetic term of electron densities, $H_{\rm pv}$ and $H_{\rm pp}$ are plane-volume and plane-plane interactions
and the superscripts (1),(2) and (v)
stand for the first plane, the second plane and volume respectively. $H_0$ and $H_{\rm ee}$ are as given in Eq.\ref{eq9}
in the first section. Remaining two interactions, $H_{\rm pv}$ and $H_{\rm pp}$, must be calculated.
\begin{eqnarray}
\nonumber H_{\rm pv}&=&\int dz\int d^2\rho\,d^2\rho'\,V(\vro-\vro\,',z)\phi^\dg(\vx)\phi(\vx)\psi^\dg(\vro\,')\psi(\vro\,') \\
\label{eq50} &&+\int dz\int d^2\rho\,d^2\rho''\,V(\vro-\vro\,'',z-d)\phi^\dg(\vx)\phi(\vx)\xi^\dg(\vro\,'')\xi(\vro\,'')
\end{eqnarray}
where $\psi$ and $\xi$ are the 2D electron field operators for the first and the second planes respectively, $\phi$ is the field operator
for the bulk electron. Using the 2D electron field operators,
\begin{subequations}
\begin{eqnarray}
\label{eq51a}\psi(\vro)=\int {d^2k\over (2\pi)^2}\,e^{i\vk\cdot\vro}\,a_{\vk}\\
\label{eq51b}\xi(\vro)=\int {d^2k\over (2\pi)^2}\,e^{i\vk\cdot\vro}\,c_{\vk}\,.
\end{eqnarray}
\end{subequations}
One gets the plane-volume interaction term, $H_{\rm pv}$, in terms of the single particle operators,
\begin{equation}\label{eq52}
H_{\rm pv}=\int {d^2k\over(2\pi)^2}{d^3p\over (2\pi)^3}{d^3p'\over (2\pi)^3}
\left[U(\vp)b^\dagger_{\vp\,'-\vp}b_{\vp\,'}a^\dagger_{\vk+\vp_\perp}a_{\vk}+U(\vp)e^{ip_\perp d}
b^\dagger_{\vp\,'-\vp}b_{\vp\,'}c^\dagger_{\vk+\vp_\perp}c_{\vk}\right].
\end{equation}
Derivation of Eq.\ref{eq52} is given in Appendix A. Using the definitions Eq.\ref{eq36} and Eq.{\ref{eq37},
one obtains for $H_{\rm pv}$,
\begin{equation}\label{eq53}
H_{\rm pv}=\int{d^3p\over (2\pi)^3}\left[U(\vp)\,B^\dg_{\vp}A_{p_\perp}+U(\vp)\,e^{ip_\perp d}B^\dg_{\vp}C_{p_\perp}\right]\,.
\end{equation}
The expression in Eq.\ref{eq53} is further reduced by integrating over the normal component $p_\perp$,
\begin{equation}\label{eq54}
H_{\rm pv}=\int{d^2k\over (2\pi)^2}\left[m^\dg_{1\vk}\,A_{\vk}+m^\dg_{2\vk}\,C_{\vk}\right]
\end{equation}
where
\begin{equation}\label{eq55}
m_{i\vk}=\int {dp_\perp\over 2\pi}(\delta_{i1}+\delta_{i2}e^{ip_\perp d}) U(\vk,p_\perp)\,B_{\vk,p_\perp}\,.
\end{equation}

The plane-plane interaction term, $H_{\rm pp}$, given by,
\begin{equation}\label{eq56}
H_{\rm pp}=\int d^2\rho_1\,d^2\rho_2\,V_{pp}(\vro_1-\vro_2,d)\,\psi^\dg(\vro_1)\psi(\vro_1)\xi^\dg(\vro_2)\xi(\vro_2)\,,
\end{equation}
similarly becomes,
\begin{eqnarray}
\nonumber H_{\rm pp}&=&\int{d^2k_1\over (2\pi)^2}{d^2k_2\over (2\pi)^2}{d^2q\over (2\pi)^2}
V_{\rm pp}(\vq,d)\,a^\dg_{\vk_1+\vq}a_{\vk_1}c^\dg_{\vk_2-\vq}c_{\vk_2} \\[1mm]
\label{eq57}&=&\int d^2q\,V_{\rm pp}(\vq,d)\,A_{\vq}C^\dg_{\vq}
\end{eqnarray}
where the plane-plane interaction $V_{\rm pp}$,
\begin{eqnarray}
\nonumber V_{\rm pp}(\vq,z)&=&e^2\int_0^{k_c}d\rho\,{\rho\over\sqrt{\rho^2+d^2}}\int_0^{2pi} d\theta\,e^{iq\rho\cos\theta} \\[1mm]
\label{eq58} &=& 2\pi e^2\int_0^{k_c} d\rho {\rho\over\sqrt{\rho^2+d^2}}\, J_0(q\rho).
\end{eqnarray}
in the limit $k_c\rightarrow\infty$ turns out to be,
\begin{equation}\label{eq59}
V_{\rm pp}(q,d)={2\pi e^2\over q}e^{-qd}=T^2_q=M^2_qe^{-qd}\,.
\end{equation}

Now, the total Hamiltonian is,
\begin{eqnarray}
\nonumber H&=& \int{d^2k\over (2\pi)^2}\Big[\varepsilon(\vk)\Big(a^\dg_{\vk}a_{\vk}+c^\dg_{\vk}c_{\vk}\Big)+
{1\over 2}M_k^2\Big(A^\dg_{\vk}A_{\vk}+C^\dg_{\vk}C_{\vk}\Big)+{1\over 2}\Big(m_{1\vk}A_{\vk}+A^\dg_{\vk}m_{1\vk}+
m_{2\vk}C_{\vk}+C^\dg_{\vk}m_{2\vk}\Big) \\[1mm]
\nonumber &&+{1\over 2}T_k^2\Big(A^\dg_{\vk}C_{\vk}+C^\dg_{\vk}A_{\vk}\Big)\Big]+\int{d^3p\over (2\pi)^3}
\,\varepsilon(\vp)\,b^\dg_{\vp}b_{\vp} \\[1mm]
\label{eq60} &=& H_0+{1\over 2}\int {d^2k\over (2\pi)^2}\Big( M_1^2A_1^\dg A_1+M_2^2A_2^\dg A_2
-{1\over M_1^2}W^\dg_1W_1-{1\over M_2^2}W^\dg_2W_2\Big)
\end{eqnarray}
where $H_0$ represents the single electron free energy terms in the two planes and in the bulk,
\[
H_0=\int{d^2k\over (2\pi)^2}\varepsilon(\vk)\Big(a^\dg_{\vk}a_{\vk}+c^\dg_{\vk}c_{\vk}\Big)+
\int{d^3p\over (2\pi)^3}\,\varepsilon(\vp)\,b^\dg_{\vp}b_{\vp}
\]
and
\begin{eqnarray}
\nonumber&&W_1={1\over\sqrt{2}}(m_1+m_2)\\
\nonumber&&W_2={1\over\sqrt{2}}(m_1-m_2)\\
\nonumber&&A_1={1\over\sqrt{2}}(A+C)+{1\over M_1^2}W_1\\
\nonumber&&A_2={1\over\sqrt{2}}(A-C)+{1\over M_2^2}W_2.
\end{eqnarray}

The above Hamiltonian is extended as before by introducing the conjugate momenta $P_1$ and $P_2$ for the two planes,
\begin{equation}\label{eq61}
H=H_0+H_c
\end{equation}
where
\begin{eqnarray}
\nonumber H_c = &&{1\over 2}\int {d^2k\over (2\pi)^2}\Big[ M_1^2A_1^\dg A_1+M_2^2A_2^\dg A_2
-{1\over M_1^2}W^\dg_1W_1-{1\over M_2^2}W^\dg_2W_2\\
\label{eq62}&&+P_1^\dg P_1+P_2^\dg P_2+M_1\Big(A_1^\dg P_1+P_1^\dg A_1\Big)+M_2\Big(A_2^\dg P_2+P_2^\dg A_2\Big)\Big].
\end{eqnarray}
After employing the Bohm-Pines transformation,
\begin{equation}\label{eq63}
S=\int_{k<k_c}{d^2k\over (2\pi)^2}\left(M_{1k}Q_{1k}A_1+M_{2k}Q_{2k}A_2\right)\,,
\end{equation}
the transformed free Hamiltonian terms are obtained as,
\begin{eqnarray}
\nonumber H_0^{(i)'} &=& \int {d^2k\over (2\pi)^2} \epsilon(\vk) (\delta_{i,1}a_{\vk}^\dg a_{\vk}+
\delta_{i,2}c_{\vk}^\dg c_{\vk}) \\ \nonumber && + i\int {d^2q\over (2\pi)^2} M_q Q_{\vq} \int {d^2k\over (2\pi)^2}
{\vq\cdot\left( \vk+{\vq\over 2}\right)\over m}\,(\delta_{i,1}a_{\vk+\vq}^\dg a_{\vk}+\delta_{i,2}c_{\vk}^\dg c_{\vk})
\\ \label{eq64} && - \int {d^2q_1\over (2\pi)^2} {d^2q_2\over (2\pi)^2} M_{q_1} M_{q_2} Q_{\vq_1} Q_{\vq_2} {\vq_1\cdot\vq_2\over m}
(\delta_{i,1}A_{\vq_1+\vq_2}+\delta_{i,2}C_{\vq_1+\vq_2})\hskip2cm(i=1,2)\\
\label{eq65} H_0^{({\rm v})'}&\cong& H_0^{(\rm v)}+i\int{d^3p\over(2\pi)^3}\int{d^3t\over(2\pi)^3}{\vt\cdot\left(\vp+{\vt\over 2}\right)
\over \sqrt{2}m}\,U(\vt)\left[{1+e^{-it_{\perp}d}\over M_{1t_\perp}}Q_{1t_\perp}+
{1-e^{-it_{\perp}d}\over M_{2t_\perp}}Q_{2t_\perp}\right]b_{\vp+\vt\,}^\dg b_{\vp}\,.
\end{eqnarray}
(For the derivation of $H_0^{({\rm v})'}$, see Appendix B.) However, we are interested in the transformed interaction
Hamiltonian $H_c'$,
\begin{equation}\label{eq66}
H_c'={1\over 2}\int{d^2k\over (2\pi)^2}\Big[P_{1,\vk}^\dg P_{1,\vk}+P^\dg_{2,\vk} P_{2,\vk}
-{1\over M_1^2}W^\dg_{1,\vk}W_{1,\vk}-{1\over M_2^2}W^\dg_{2,\vk}W_{2,\vk}\Big]\,.
\end{equation}

Since it provides the interaction Hamiltonian for the bulk electron-plane system,
\begin{eqnarray}
\nonumber H_{\rm eff}&=&-{1\over 2}\int {d^2k\over (2\pi)^2} \left[{1\over M_{1k}^2}W^\dg_{1k}W_{1k}+
{1\over M_{2k}^2}W^\dg_{2k}W_{2k}\right]\\[2mm]
\nonumber &=&-{1\over 4}\int{d^2k\over (2\pi)^2}\left[\left(m^\dg_{1k}m_{1k}+m^\dg_{2k}m_{2k}\right)
\left({1\over M_{1k}}+{1\over M_{2k}}\right)\right.\\ \label{eq67}
&&\hspace{18mm}+\left.\left(m^\dg_{1k}m_{2k}+m^\dg_{2k}m_{1k}\right)
\left({1\over M_{1k}}-{1\over M_{2k}}\right)\right]\,,
\end{eqnarray}
and becomes,
\begin{eqnarray}
\nonumber H_{\rm eff}&=&-{1\over 2}\int {d^2k\over (2\pi)^2}\left\{{1\over \widetilde{V}(k)\left(1-e^{-2kd}\right)}
\int{d\ell\over 2\pi}{d\ell'\over 2\pi}U(\vk,\ell)U(\vk,\ell')\left[1+e^{i(\ell'-\ell)d}\right]\right.\\
\label{eq68} &&\hspace{19mm}-\left.{e^{-kd}\over \widetilde{V}(k)\left(1-e^{-2kd}\right)}
\int{d\ell\over 2\pi}{d\ell'\over 2\pi}U(\vk,\ell)U(\vk,\ell')\left[e^{i\ell'd}+e^{-i\ell'd}\right]\right\}
B^\dg_{\vk,\ell'}B_{\vk,\ell}\,.
\end{eqnarray}

The effective potential energy $V_{\rm eff}$ for this system is read out as done in Eq.\ref{eq45b},
\begin{eqnarray}
\nonumber V_{\rm eff}(\vx,\vx',k_c)&=&-\int {d^2k\over (2\pi)^2}e^{i\vk\cdot(\vro-\vro\,')}\widetilde{V}(k)^{-1}(1-e^{-2kd})^{-1}
\\ \nonumber &&\times \int{d\ell\over 2\pi}{d\ell'\over 2\pi}U(\vk,\ell)U(\vk,\ell')\Bigg\{\Big[1+e^{i(\ell'-\ell)d}\Big]
-e^{-kd}\Big[e^{i\ell'd}+e^{-i\ell'd}\Big]\Bigg\}e^{i(\ell z-\ell'z')}\,.\\ \label{eq69}
\end{eqnarray}

Evaluation of $V_{\rm eff}$ involves integrals of the form $\int{d\ell\over 2\pi}U(\vk,\ell)e^{i\alpha\ell}$ where
$U(\vk,\ell)=4\pi e^2/(k^2+\ell^2)$. The result of $\ell$-integration is $\widetilde{V}(k) e^{-|\alpha|k}$ where
$\widetilde{V}(k)=2\pi e^2/k$. The remaining 2D integrals give,

\begin{eqnarray}
\nonumber V_{\rm eff}(\vx,\vx',k_c)&=&-\int {d^2k\over (2\pi)^2}e^{i\vk\cdot(\vro-\vro\,')}{V(k)\over 1-e^{-2kd}}
\Bigg\{\left[e^{-k(|z|+|z'|)}+e^{-k(|z-d|+|z'-d|)}\right]\\ \label{eq70}
&&\hspace{39mm}-e^{-kd}\left[e^{-k(|z|+|z'-d|)}+e^{-k(|z-d|+|z'|)}\right]\Bigg\}\,.
\end{eqnarray}

The terms in Eq.\ref{eq70} are evaluated by using the result,
\begin{eqnarray}
\nonumber-\int{d^2k\over(2\pi)^2}e^{i\vk\cdot\Delta}{\widetilde{V}(k)\over 1-e^{-2kd}}e^{-sk}&=&
-e^2\int_0^{\pi}{d\theta\over 2\pi}\int_0^{k_c} {e^{k(-s+i\Delta\cos\theta)}\over 1-e^{-2kd}}\\
\nonumber&=&-e^2\int_0^{k_c} dk{e^{-ks}\over 1-e^{-2kd}}J_0(k\Delta)\,.\\[3mm]
\noalign{\rm The $k_c\to\infty$ limit gives,}\nonumber \\[1mm]
\nonumber&=&\sum_{n=0}^\infty\int_0^\infty dk e^{-k(s+2nd)}J_0(k\Delta)\\
\nonumber&=&\sum_0^\infty\left[\Delta^2+(s+2nd)^2\right]^{-1/2}
\end{eqnarray}
and one finally obtains the effective potential to be a sum of interactions between infinite number of images,
\begin{eqnarray}
\nonumber V_{\rm eff}(\vx,\vx\,')&=&-e^2\sum_{n=0}^\infty\left\{\left[|\vro-\vro\,'|^2+(|z|+|z'|+2nd)^2\right]^{-1/2}
+\left[|\vro-\vro\,'|^2+(|z-d|+|z'-d|+2nd)^2\right]^{-1/2}\right.\\[2mm]
\nonumber &&-\left.\left[|\vro-\vro\,'|^2+(|z|+|z'-d|+(2n+1)d)^2\right]^{-1/2}+
\left[|\vro-\vro\,'|^2+(|z-d|+|z'|+(2n+1)d)^2\right]^{-1/2}\right\}\\
\label{eq71}
\end{eqnarray}

For the case of a single bulk charge at $\vx=\vx\,'$ one obtains the result for the bulk point charge between two
parallel planes in terms of infinite number
of images.
\begin{eqnarray}
\nonumber V_{\rm eff}(z)&=&-{e^2\over 2}\sum_n\left[{1\over nd+z}+{1\over nd+(d-z)}-{2\over (n+1)d}\right]\\
\nonumber &=&-{e^2\over 2d}\sum_n \left({1\over n+{z\over d}}+{1\over n+(1-{z\over d})}-{2\over n+1}\right)\\
\nonumber &=&-{e^2\over 2d}\sum_n{1\over (n+{z\over d})(n+1-{z\over d})}-{e^2z\over d^2}\Big[{z\over d}-1\Big]
\sum_n{1\over (n+1)(n+{z\over d})(n+1-{z\over d})}
\end{eqnarray}
Setting the position to $z=d/2$, the result is simplified as,
\begin{eqnarray}
\nonumber V_{\rm eff}&=&-{e^2\over 2d}\sum_n{1\over (n+1)(n+{1\over 2})}\\
\nonumber &=&-{e^2\over 2d}\left[2+2\int_0^1 {x^{1/2}-x\over 1-x}dx\right]\\
\nonumber &=&-{e^2\over d}\ln 2
\end{eqnarray}

\section{Conclusion}

The application of Bohm-Pines transformation to the electron-conducting surface problem has resulted in an effective
two-particle interaction potential for the system. The transformed Hamiltonian contains the screened interaction
($k<k_c$) as well as the long range part of the Coulomb interaction ($k>k_c$). The two-point interaction
term explicitly demonstrates the quantum dynamics of the electron with an image electron, which reduces to the
well known classical image method results. The transformation appears to be a powerful method to study more complicated
surface geometries.

\appendix
\section{}
\begin{eqnarray}\label{app1}
\nonumber H_{\rm pv}&=&\int d^2\rho d^2\rho'{d^3p_1\over (2\pi)^3}{d^2k'_1\over(2\pi)^2}{d^2k'_2\over(2\pi)^2}{d^3p_2\over (2\pi)^3}
e^{i(\vk_2-\vk_1)\cdot\vro}e^{i(\vk_2'-\vk_1')\cdot\vro\,'}b^\dagger_{\vp_1}a^\dagger_{\vk_1'}a_{\vk_2'}b_{\vp_2}
\int dz e^{i(\ell_2-\ell_1)z} V(\vro-\vro\,',z)\\ \nonumber
&&+\int d^2\rho d^2\rho''{d^3p_1\over (2\pi)^3}{d^2k''_1\over(2\pi)^2}{d^2k''_2\over(2\pi)^2}{d^3p_2\over (2\pi)^3}
e^{i(\vk_2-\vk_1)\cdot\vro}e^{i(\vk_2''-\vk_1'')\cdot\vro\,''}b^\dagger_{\vp_1}a^\dagger_{\vk_1''}a_{\vk_2''}b_{\vp_2}
\int dz e^{i(\ell_2-\ell_1)z} V(\vro-\vro\,'',z-d)\\ \nonumber
&=&\int{d^3p_1\over (2\pi)^3}{d^2k'_1\over(2\pi)^2}{d^2k'_2\over(2\pi)^2}{d^3p_2\over (2\pi)^3}
b^\dagger_{\vp_1}a^\dagger_{\vk_1'}a_{\vk_2'}b_{\vp_2}\int d^2\rho d^2\rho'e^{i(\vk_2-\vk_1)\cdot\vro}
e^{i(\vk_2'-\vk_1')\cdot\vro\,'}\widetilde{V}(\vro-\vro\,',\ell_2-\ell_1)\\ \nonumber
&&+\int{d^3p_1\over (2\pi)^3}{d^2k''_1\over(2\pi)^2}{d^2k''_2\over(2\pi)^2}{d^3p_2\over (2\pi)^3}
b^\dagger_{\vp_1}c^\dagger_{\vk_1''}c_{\vk_2''}b_{\vp_2}\int d^2\rho d^2\rho'e^{i(\vk_2-\vk_1)\cdot\vro}
e^{i(\vk_2''-\vk_1'')\cdot\vro\,''}e^{i(\ell_2-\ell_1)d}\widetilde{V}(\vro-\vro\,'',\ell_2-\ell_1)\\
\end{eqnarray}
Substitution $\vec r_1=\vro\,'$, $\vec r_2=\vro-\vro\,'$ in the first integral and $\vec r_1=\vro\,''$, $\vec r_2=\vro-\vro\,''$
in the second one gives
\begin{eqnarray}\label{app2}
\nonumber H_{\rm pv}&=&\int{d^3p_1\over (2\pi)^3}{d^2k'_1\over(2\pi)^2}{d^2k'_2\over(2\pi)^2}{d^3p_2\over (2\pi)^3}
b^\dagger_{\vp_1}a^\dagger_{\vk_1'}a_{\vk_2'}b_{\vp_2}\int d^2r_2\widetilde{V}(\vec r_2,\ell_2-\ell_1)
e^{i(\vk_2-\vk_1)\cdot\vec r_2}\int d^2r_1 e^{i[(\vk_2-\vk_1)+(\vk_2'-\vk_1')]\cdot\vec r_1}\\ \nonumber
&&+\int{d^3p_1\over (2\pi)^3}{d^2k''_1\over(2\pi)^2}{d^2k''_2\over(2\pi)^2}{d^3p_2\over (2\pi)^3}
b^\dagger_{\vp_1}c^\dagger_{\vk_1''}c_{\vk_2''}b_{\vp_2}e^{i(\ell_2-\ell_1)d}
\int d^2r_2\widetilde{V}(\vec r_2,\ell_2-\ell_1)
e^{i(\vk_2-\vk_1)\cdot\vec r_2}\int d^2r_1 e^{i[(\vk_2-\vk_1)+(\vk_2''-\vk_1'')]\cdot\vec r_1}\\ \nonumber
&=&\int{d^3p_1\over (2\pi)^3}{d^3p_2\over (2\pi)^3}{d^2k'_1\over(2\pi)^2}d^2k_2' U(\vk_2-\vk_1,\ell_2-\ell_1)
\delta(\vk_2-\vk_1+\vk_2'-\vk_1')b^\dagger_{\vp_1}a^\dagger_{\vk_1'}a_{\vk_2'}b_{\vp_2}\\ \nonumber
&&+\int{d^3p_1\over (2\pi)^3}{d^3p_2\over (2\pi)^3}{d^2k''_1\over(2\pi)^2}d^2k_2'' e^{i(\ell_2-\ell_1)d}
U(\vk_2-\vk_1,\ell_2-\ell_1)\delta(\vk_2-\vk_1+\vk_2''-\vk_1'')b^\dagger_{\vp_1}c^\dagger_{\vk_1''}c_{\vk_2''}b_{\vp_2}\\ \nonumber
&=&\int{d^3p_1\over (2\pi)^3}{d^3p_2\over (2\pi)^3}{d^2k'_1\over(2\pi)^2}U(\vp_2-\vp_1)
b^\dagger_{\vp_1}b_{\vp_2}a^\dagger_{\vk_1'}a_{\vk_1+\vk_1'-\vk_2}\\ \nonumber
&&+\int{d^3p_1\over (2\pi)^3}{d^3p_2\over (2\pi)^3}{d^2k''_1\over(2\pi)^2}e^{i(\ell_2-\ell_1)d}
U(\vp_2-\vp_1)b^\dagger_{\vp_1}b_{\vp_2}c^\dagger_{\vk_1''}c_{\vk_1+\vk_1''-\vk_2}\\
\end{eqnarray}
We substitute $\vp\,'=\vp_2$ and $\vp=\vp_2-\vp_1$ to get,
\begin{equation}\label{app3}
H_{\rm pv}=\int {d^2k\over(2\pi)^2}{d^3p\over (2\pi)^3}{d^3p'\over (2\pi)^3}
\left[U(\vp)b^\dagger_{\vp\,'-\vp}b_{\vp\,'}a^\dagger_{\vk+\vp_\perp}a_{\vk}+U(\vp)e^{ip_\perp d}
b^\dagger_{\vp\,'-\vp}b_{\vp\,'}c^\dagger_{\vk+\vp_\perp}c_{\vk}\right]
\end{equation}

\section{}
\begin{equation}\label{app4}
[S,H_0^{(\rm v)}]=\int{d^2k\over (2\pi)^2}\int{d^3p\over (2\pi)^3}\,\varepsilon_p\Big\{M_{1k}Q_{1,\vk}
[A_{1,\vk},b^\dg_{\vp}b_{\vp}]+M_{2k}Q_{2,\vk}[A_{2,\vk},b^\dg_{\vp}b_{\vp}]\Big\}
\end{equation}
where
\begin{equation}\label{app5}
[A_{i,\vk},b^\dg_{\vp}b_{\vp}]={1\over \sqrt{2}M_{ik}^2}\left\{[m_{1k},b^\dg_{\vp}b_{\vp}]\pm
[m_{2k},b^\dg_{\vp}b_{\vp}]\right\}\hskip5mm i=1,2
\end{equation}
and
\begin{eqnarray}
\nonumber [m_{ik},b^\dg_{\vp}b_{\vp}]&=&\int {dt_\perp\over 2\pi} U(\vt)\big(\delta_{i1}+\delta_{i2}e^{-ip_\perp d}\big)
\big[B_t,b^\dg_{\vp}b_{\vp}\big]_{t_\perp=k}\\[2mm]
\label{app6}&=& -\int{dt_\perp\over 2\pi} U(\vt)\big(\delta_{i1}+\delta_{i2}e^{-ip_\perp d}\big)
\left[b^\dg_{\vp-\vt\,}b_{\vp}-b^\dg_{\vp+\vt\,}b_{\vp}\right]_{t_\perp=k}
\end{eqnarray}

Then
\begin{eqnarray}
\label{app7}&&\int{d^3p\over (2\pi)^3}\varepsilon_p[A_{1,\vk},b^\dg_{\vp}b_{\vp}]=-{1\over\sqrt{2}M^2_{1k}}
\int{d^3p\over (2\pi)^3}\int {dt_\perp\over 2\pi} U(\vk,t_\perp)\left(1+e^{-it_\perp d}\right)
\left.(\varepsilon_{p+t}-\varepsilon_p)b^\dg_{\vp+\vt\,}b_{\vp}\right\vert_{t_\perp=k}\\[2mm]
\label{app8}&&\int{d^3p\over (2\pi)^3}\varepsilon_p[A_{2,\vk},b^\dg_{\vp}b_{\vp}]=-{1\over\sqrt{2}M^2_{2k}}
\int{d^3p\over (2\pi)^3}\int {dt_\perp\over 2\pi} U(\vk,t_\perp)\left(1-e^{-it_\perp d}\right)
\left.(\varepsilon_{p+t}-\varepsilon_p)b^\dg_{\vp+\vt\,}b_{\vp}\right\vert_{t_\perp=k}
\end{eqnarray}

Therefore, transformation of kinetic term for the volume can be written as,
\begin{equation}\label{app9}
H_0^{(\rm v)'}\cong H_0^{(\rm v)}+i\int{d^3p\over(2\pi)^3}\int{d^3t\over(2\pi)^3}{\vt\cdot\left(\vp+{\vt\over 2}\right)
\over \sqrt{2}m}\,U(\vt)\left[{1+e^{-it_{\perp}d}\over M_{1t_\perp}}Q_{1t_\perp}+
{1-e^{-it_{\perp}d}\over M_{2t_\perp}}Q_{2t_\perp}\right]b_{\vp+\vt\,}^\dg b_{\vp}.
\end{equation}

\end{document}